\documentclass{article}
\usepackage{PRIMEarxiv}
\usepackage{amsmath}
\usepackage[utf8]{inputenc} % allow utf-8 input
\usepackage[T1]{fontenc}    % use 8-bit T1 fonts
\usepackage{hyperref}       % hyperlinks
\usepackage{url}            % simple URL typesetting
\usepackage{booktabs}       % professional-quality tables
\usepackage{amsfonts}       % blackboard math symbols
\usepackage{nicefrac}       % compact symbols for 1/2, etc.
\usepackage{microtype}      % microtypography
\usepackage{lipsum}
\usepackage{fancyhdr}       % header
\usepackage{graphicx}       % graphics
\usepackage{xcolor}
\graphicspath{{media/}}     % organize your images and other figures under media/ folder
\usepackage{multicol}
\usepackage{longtable}
\usepackage{chemformula}
\title{REDUCING CARTEL RECRUITMENT IS THE ONLY WAY TO LOWER VIOLENCE IN MEXICO}

\author{
  Rafael Prieto-Curiel\\
  Complexity Science Hub \\ Josefstaedter Strasse 39 \\ 1080 Vienna, Austria \\
  \texttt{prieto-curiel@csh.ac.at} \\
   \And
  Gian Maria Campedelli \\
  University of Trento \\ Trento \\ Italy\\
  \texttt{gianmaria.campedelli@unitn.it}\\
  \And
  Alejandro Hope\\
  Independent Security Analyst \\ Mexico City \\ Mexico
}

\begin{document}
\maketitle

\begin{abstract}
Every year, Mexican cartels lose many members due to conflict with other cartels and arrests. Yet, despite their losses, cartels have managed to increase violence for years. We address this puzzle by leveraging data on the number of homicides, missing persons and arrests in Mexico for the past ten years, along with information on the interactions among cartels aiming to estimate the size of cartels' population. We model recruitment, incapacitation by the state, conflict and saturation as the reasons why cartels vary in size. Results show that by 2022 cartels have between 160,000 and 185,000 units, becoming one of the top employers in the country. Recruiting at least 350 people per week is essential to avoid their collapse due to the aggregate effect of incapacitation, conflict and saturation. Furthermore, we test the effects of two policy scenarios aimed at decreasing cartel violence. We show that given the state of violence, increasing current levels of incapacitation leads to a rise in homicides and cartel members. Conversely, reducing recruitment provides substantial benefits in terms of violence reduction and decreasing cartel population, calling for structural investments in a proactive strategy that targets individuals at risk rather than a traditional reactive approach centred around incapacitation.

\end{abstract}

\section{Introduction}

%%% cartels are super violent
{
Latin America is home to only 8\% of the world's population, but roughly one in three intentional homicides worldwide occur in the region \cite{UNODC}. Mexico accounts for a relevant share of homicides in the region, especially due to the longstanding presence of cartels across many areas of the country. In 2021 Mexico reported 34,000 victims of intentional homicide, nearly 27 victims per 100,000 inhabitants and was ranked among the least peaceful countries in Latin America \cite{InstituteforEconomicsandPeaceGlobalPeaceIndex2022}. Between 2007 and 2021, the number of homicides in the country increased by more than 300\% \cite{InegiMortalidad}, with institutional sources quantifying that in the 2006-2018 window, about 125,000-150,000 homicides were organised crime-related in Mexico \cite{CongressionalResearchServiceMexicoOrganizedCrime2020}. 
}

%%% violence is not all the effect, and drugs are a minor element of cartels
{
The effects of cartels on Mexico's society are far-reaching. Over the last decades, organised crime has significantly affected society through extensive acts of violence, threatening institutional stability and people's safety, deteriorating human rights, and infiltrating institutions through the corruption of politicians and justice system members \cite{WrightNecropoliticsNarcopoliticsFemicide2011, TrellesBulletsVotesViolence2012, MorrisCorruptionDrugTrafficking2012, BlumeOldRulesNo2017, Anaya-MunozMexicoHumanRights2018}. Recent estimates attest that violence in Mexico costed 243 US\$ billion dollars in 2021 alone, corresponding to 20.8\% of the country's GDP \cite{InstituteforEconomicsandPeaceMexicoPeaceIndex2022}. Cartels are central actors in the illicit drug market but are also active in extortion, disappearances, kidnapping and other illegal activities \cite{Jonesunintendedconsequenceskingpin2013, Diaz-CayerosCaughtcrossfiregeography2015}. These organisations have been able to infiltrate different economic sectors, including the trade of wildlife, minerals and even avocados and lime \cite{Garcia-PonceHowdoesdrug2021,henkin2020pits}. Mexico's oil company (PEMEX) estimated that cartel-directed fuel theft costs the company 1.6 US\$ billion dollars a year \cite{jones2019huachicoleros}. Additionally, Mexican cartels have increasingly acquired a transnational dimension, chiefly through drug trafficking, establishing partnerships with other organised crime groups in other countries, such as Italy and the Netherlands, and expanding their businesses in adjoining countries, including the United States \cite{RubinoSinaloacartelmove2020, PopMexicanCartelsAre2020, ShuklaevolvingproblemMethamphetamine2012, MedelMexicodrugnetworks2015}.  
}

%%% cartels are black boxes
{
Yet, despite Mexican drug cartels' economic, social, and political importance, we lack essential information to understand better –-- empirically and beyond --– how they function. In fact, we lack estimates of the size of these criminal entities. We also lack systematic estimates of cartel-related killings and kidnappings as well as figures related to recruitment trends, making it extremely difficult to deepen our knowledge about their presence, resources, and goals. The secretive nature of cartels' actions, as well as the insufficient amount of information accessible to map them, makes them conceptually similar to black boxes, from which we can only extrapolate imperfect proxies of activity, using, for instance, the daily number of homicides or the number of drug-related arrests occurred in the country \cite{jones2022mexico}. Although homicide and arrest trends are imperfect because they do not discriminate between offences that occurred specifically in the context of organised crime, they can be used to estimate cartels' violence capacity and the state's incapacitation against them. Here, we build on this intuition and exploit data on murders, missing persons, and arrests in Mexico between 2012 and 2022 to derive cartel size, proposing a mathematical system to represent their behaviour over ten years and seeking to shed light on the mechanisms within cartels' black box. 
}

%%% we use data for murders and arrests
{
The present work has two main goals. First, it aims to obtain plausible estimates of the cartels' population, including their number of members and recruitment capacity. Second, it seeks to simulate different policy scenarios (i.e., increased state incapacitation and recruitment prevention) to disentangle the effects of varying strategies to curb cartels' power and, in turn, violence in the country.
}
%%%  model recruitment, incapacitation, ...

{
Our conceptual framework is built on the evidence that, despite the high number of murders and arrests in the last ten years, cartels have maintained and even increased their power, control, and resources, introducing even more violence in the country. While cartels lose dozens of members daily due to killings and state incapacitation through arrest, the violence over the years has not decreased. We tackle this puzzle by studying cartels' evolution, deriving their size, and considering four fundamental sources of size variation: recruitment, incapacitation, conflict, and saturation. These sources capture the different exogenous and endogenous dynamics explaining why and to what extent cartels grow or shrink. Recruitment refers to the process of attracting new workforce which stably carries out tasks (both strictly criminal and not) for cartels \cite{CalderoniOrganizedcrimegroups2022}. Incapacitation measures the ability of the state to counter cartels through arrests \cite{CohenIncapacitationStrategyCrime1983}. Considering all arrests allows us to avoid the bias of only focusing on arrests for homicides, which are only a fraction of the offences committed by cartel members. Conflict describes the extent to which cartels clash and fight with each other \cite{TrejoWhyDidDrug2018}. Finally, saturation characterises internal instability and dropouts, leading to organisational fragmentation \cite{AtuestaFragmentationcooperationevolution2018a, CongressionalResearchServiceMexicoOrganizedCrime2020}.  
}

%%% our big contribution is
{
In the past two decades, agent-based modelling and statistical simulations have gained momentum in the study of complex criminal phenomena \cite{d2015statistical,campedelli2022crimeml}. Among other topics, scholars have computationally addressed the study of recruitment in criminal organisations, mafias' protection racketeering, and radicalisation, as well as the impact of deterrence and institutional interventions to counter crime \cite{calderoni2022recruitment, NardinGLODERSSsimulatoragentbased2017a, chuang2019mathematical, feichtinger2002optimal, caulkins2008optimizing, caulkins2009optimal}. At a high level, we contribute to this evolving strand of research by highlighting the potential that mathematical approaches can have for analysing complex criminal phenomena without relevant data. To construct our model, we gauge data on 150 cartels active in Mexico in 2020, including information on their alliances and rivalries and data corresponding to homicides, missing persons and arrests. We estimate that cartels are the fifth largest employer in Mexico. Furthermore, we estimate that unless all cartels combined recruited at least 350 people per week, they would have collapsed due to the aggregate effect of conflict, incapacitation and saturation. Despite massive cartel losses, the increasing violence marking the last years is chiefly driven by recruitment. Recruitment acts as a compensating process, preventing cartels from vanishing due to fierce conflicts with other criminal organisations and the state. We finally show that even doubling the current levels of incapacitation, with all the challenges this process would involve, will result in an increase in the weekly number of casualties by more than 8\%. Conversely, reducing cartel recruitment by half would reduce casualties in the country by nearly 25\%. 
}

\section{Results}

\subsection{Estimating Cartels' Population}

%%%% PRESENT MODEL. Figure 1
{
Most cartel-related activities are organised as dark networks, often with transnational characteristics, to maintain their operations and activities covered \cite{bakker2012preliminary, raab2003dark, martins2022universality, wachs2019network}. However, their human losses caused by homicidal violence and the state's action via incapacitation provide insights into the overall amount of such activities. We leverage the trends in homicides, missing persons, and arrests over the last decade to motivate our investigation of cartels' size in Mexico (Supplementary information A). Not all losses are directly related to the conflict between cartels (for example, domestic violence), and some are a byproduct of their dispute (for example, deaths suffered by family members or bystanders). To study the size and evolution of the cartel population, we only model homicides between cartel members. Albeit only a fraction of the total number of homicides and arrests in the country are suffered directly by cartel members, cartels have not seen their power diminished since violence has not reduced either. In Mexico, 686 people were killed each week of 2021, with an additional 137 people reported as missing and yet to be found, and more than 2,500 people imprisoned each week \cite{InegiMortalidad, RNPDNO, inegiPenitenciario}. 
}

%%%% losses are stable and we use them figure 2
{
We use the number of cartel losses to infer otherwise unknown properties, including their size and recruitment rate. Data compiled from open sources in Mexico \cite{BacrimMexico} enable us to detect the existence of $\kappa = 150$ active cartels in Mexico in 2020 (Supplementary information B). Cartels have different interactions: they can be allies or have no interactions (particularly from distant locations), or they can fight for territory or resources across multiple locations, creating significant losses among both groups. To represent these interdependencies, we construct two separate weighted networks: the allies $\mathcal{A}$ and rivalries $\mathcal{R}$ to recreate conflicting and cooperating cartels, with weights corresponding to the number of states in which two cartels interact (Figure \ref{RecruitmentFigureNetwork}). Major cartels, like \textit{Cártel Jalisco Nueva Generación} (CJNG), the Sinaloa Cartel and Nueva Familia Michoacana are present almost at a national level and have alliances with many satellite organisations forming three main clusters. These clusters fight against each other, creating most of the violence between cartels \cite{jones2022mexico}. Smaller organisations are local to one city and tend to have few interactions (cooperation or conflict) with other cartels.  
\begin{figure}[h] \centering
\begin{center}
\includegraphics[width = 0.7\linewidth]{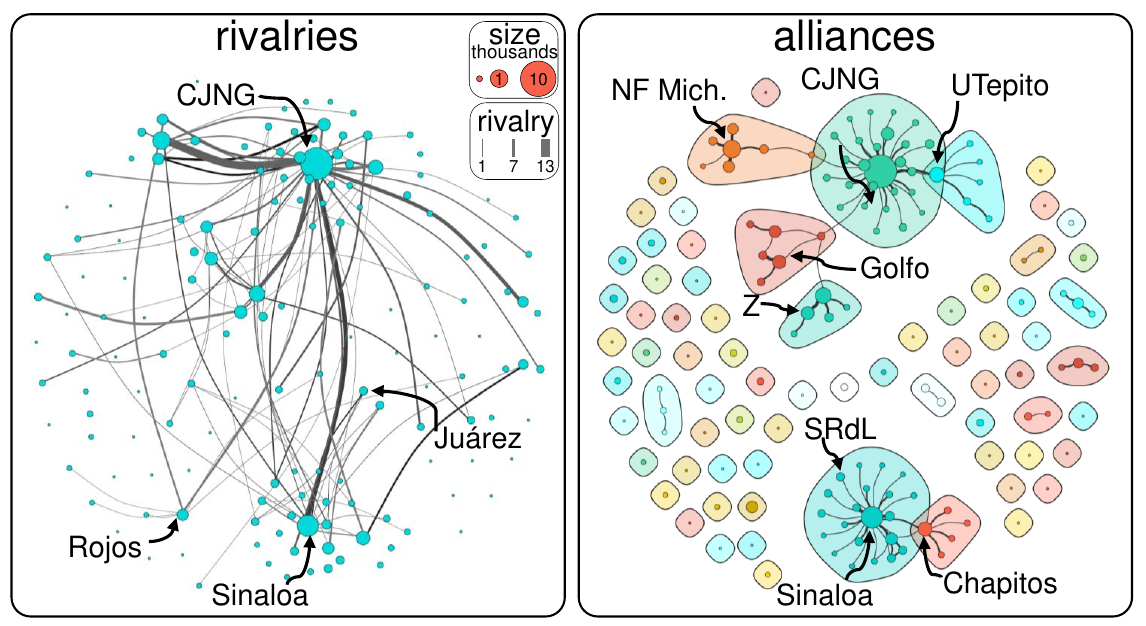}
\end{center}
\caption{Rivalries and alliances were observed between 150 active cartels in Mexico in 2020. The size of the node represents the estimated cartel size. Nodes are connected if cartels have at least one state rivalry (left). The width of the edge corresponds to the number of states in which cartels fight. Nodes are connected if they are identified as allies (right).} \label{RecruitmentFigureNetwork}
\end{figure}
}

%%% four mechanisms with equation
{
We consider four mechanisms explaining why cartel size varies: recruitment, incapacitation, saturation and conflict (Figure \ref{RecruitmentFigure1}). The number of members of cartel $i$ at time $t$, expressed as $C_i(t)$, increases instantly according to $\rho C_i$, where $\rho$ is the fixed recruitment rate. Due to state forces, the size of the cartel decreases by $\eta C_i/\sum_j C_j$ for some $\eta>0$ that represents the incapacitation rate. Due to internal instability, dropouts and diminishing returns, large groups decrease their size instantly by $\omega C_i^2$ for some small value of $\omega > 0$, known as the saturation rate \cite{coase1937nature, caulkins2009optimal}. The impact of conflict between two cartels, $i$ and $j$, is modelled according to the number of homicide offenders between rival groups, assumed to be proportional to cartel size, so cartel $i$ suffers instant casualties according to $\theta C_i C_j$, where $\theta \geq 0$ is the deathly rate of conflict related to homicide offenders within cartels. Combining recruitment, incarceration, conflict and saturation, we get that
%\begin{equation} \label{MasterEquation}
%\dot{C}_i = \rho C_i - \eta \frac{C_i}{C} - \theta \sum_{j \neq i}^{\kappa} C_i C_j S_{ij} - \omega C_i^2,
%\end{equation}
\begin{equation}\label{eq:MasterEquation}
\dot{C}_i = \underbrace{\rho C_i}_\text{recruitment} - \underbrace{\eta \frac{C_i}{C}}_\text{incapacitation} - \underbrace{\theta \sum_{j \neq i}^{\kappa} C_i C_j S_{ij}}_\text{conflict} - \underbrace{\omega C_i^2}_\text{saturation}. 
\end{equation}
where $\dot{C}_i$ indicates the rate of change in cartel size $i$ and $S_{ij} \geq 0$ captures the interaction between cartels. We obtain a system of $\kappa = 150$ coupled differential equations, one for each cartel (see the Methods section). In line with previous works on other types of organisations, we assume that the initial cartel size is a heavy-tailed distribution (more details in the Methods section) \cite{clauset2009power, johnson2006universal, restrepo2020computational, johnson2005old}. We use the observed weekly number of casualties and incapacitations to estimate the time-varying number of members of each cartel $C_i(t)$.

\begin{figure}[h] \centering
\begin{center}
\includegraphics[width = 0.6\linewidth]{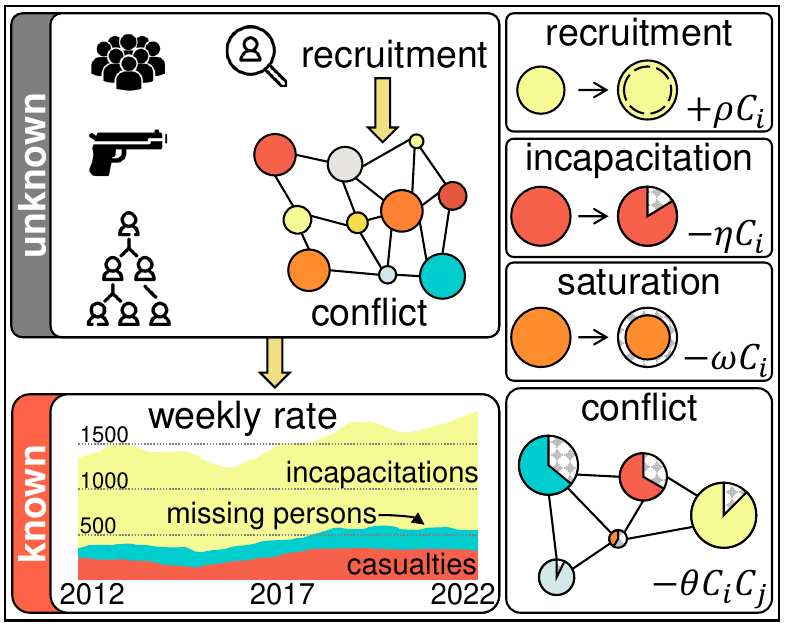}
\end{center}
\caption{Most cartel-related activities remain undercover, but we observe some of their byproducts in casualties and incapacitations. Model diagram representing the four reasons why a cartel changes in size. } \label{RecruitmentFigure1}
\end{figure}

}

%%% numbers on cartels
{
As previously mentioned, not all observed deaths, missing persons and incapacitations in the country are suffered by cartel members and also, most incapacitations are not linked to the arrest of cartel members. In our analysis, we estimate casualties as the sum of missing persons with murders and consider that a fraction $f = 10\%$ of the observed weekly deaths and a fraction $g = 5\%$ of the incapacitations are cartel members. In total, 50,000 casualties and 55,000 incapacitations are taken directly involving cartel members. Based on these figures, we estimate that in 2012 there were 115,000 cartel members and that in ten years, the number increased to 175,000. Thus, despite efforts from the state to hinder their power, cartels have increased their size by 60,000 members in a decade. Arresting nearly 6,000 cartel members each year has not prevented them from growing into larger organisations. Given the current conditions, we quantify 120 weekly cartel-related deaths, with an increase of 77\% between 2012 and 2022. To ensure that our results are not driven by wrong assumptions on the number of homicides between cartel members and incarcerations of cartels affiliates, we conduct sensitivity tests considering the scenarios between 40,000 and 60,000 cartel casualties, and 45,000 and 65,000 incapacitations. By considering the variation of these two parameters, we obtain that the total population of cartel members in 2021 lies between 160,000 and 185,000 units. At the same time, additional sensitivity tests sought to quantify the impact of potential missing data at the network level concerning alliances and rivalries. We find that adding 10\% more cartels would, on average, lead to 3.2\% more members than the estimated 175,000. Furthermore, we also provide evidence that adding 10\% more alliances or rivalries would at most impact the overall dimension of violence by 5\% (Supplementary information F). Even under a conservative scenario, Mexican cartels have lost around 200 members per week for years (Figure \ref{RecruitmentFigure2}). Specifically, we estimate that in a decade, 285,000 people acted as cartel units and that --- in total --- 38\% of them are either deceased (18\%) or incarcerated (20\%).

\begin{figure}[h] \centering
\begin{center}
\includegraphics[width = 0.7\linewidth]{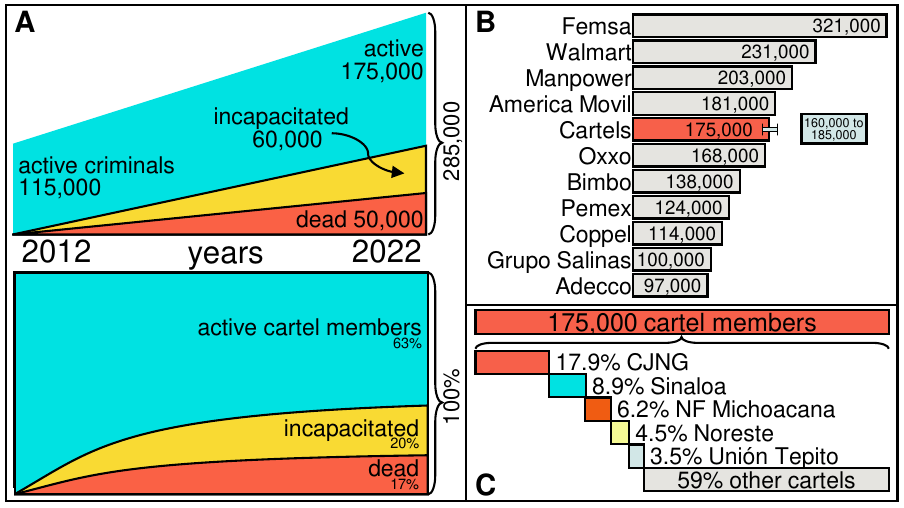}
\end{center}
\caption{A - Between 2012 and 2022, we estimate that 285,000 people took part as cartel members, but only 60\% were still active by 2022. Roughly 18\% of them are dead, and 20\% were incapacitated. B - Number of employees from the top 10 companies in Mexico and the combined size of cartels \cite{Top500Empresas}. We estimate that cartels had 175,000 members by 2022, with an interval between 160,000 and 185,000 members combined. C - Of the 175,000 active cartel members, roughly 17.9\% are part of CJNG, 8.9\% of Cartel de Sinaloa, and 6.2\% from Nueva Familia Michoacana, the top three cartels in terms of size. } \label{RecruitmentFigure2}
\end{figure}
}

%%%% numbers 2023-01
{
Despite the competition with other cartels and incapacitation by the state forces, cartels have managed to prevail for decades. These figures outline that the cartel's power lies in its recruiting capacity. Our results reveal that between January and December 2021, cartels recruited 19,300 individuals, losing 6,500 members due to conflict with other cartels and 5,700 members due to incapacitation, having a net gain of roughly 7,000 members during that year (Supplementary information C). A similar estimate is observed for each year between 2012 and 2022 (Supplementary information C). Unless all cartels combined recruit at least 370 people per week, they would have collapsed due to conflict, incapacitation and saturation combined (Figure \ref{RecruitmentFigure2}-A). However, the cartel career is brief and risky. 
}

%  conflict between the 140 smallest cartels only represents 34\% of the fatalities
{
Given the estimated overall population, all cartels combined are the fifth biggest employer in Mexico \cite{Top500Empresas} (Figure \ref{RecruitmentFigure2}-B). The ten biggest cartels in Mexico have more than 50\% of the active affiliates in the country (Figure \ref{RecruitmentFigure2}-C), but the conflict between them only produces 15\% of the fatalities. Conversely, most cartels are small local organisations playing a critical role in creating violence in the country, often becoming targets of more powerful organisations. Previous research suggests that big cartels frequently adopt fragmented cells of other weaker and less experienced structures \cite{jones2022mexico}. Small cartels play a crucial role as they are more likely to become targets of powerful illicit organisations rather than fighting organisations of similar size. We estimate that more than half of the country's casualties result from the fight between the smallest 140 and the biggest ten cartels (Supplementary information B).
}

\subsection{Comparing Policy Scenarios}
%%%% Varying recruitment and incapacitations
{
Based on the size of cartels in 2022 and the trends observed in the past decade, we also predict that the weekly number of casualties related to organised crime will keep increasing in the following years. Given the current cartel size and conflict, we estimate that if current trends continue, cartels will keep increasing their size and power, and we could observe 40\% more casualties and 26\% more cartel members by 2027 . In light of these forecasts, we test the effectiveness of two main policy scenarios designed to reduce future violence in the country. We compare a preventive strategy against organised crime, aimed at reducing cartel recruitment, against a reactive strategy, aimed at increasing incapacitation by varying the value of the related parameters. To assess which approach works best, we simulate future trends using as outcomes the corresponding number of casualties and the size of the cartels' population. On the one side, doubling incapacitation, with all the associated costs and challenges in increasing security resources (including police personnel, army, prisons, etc.), will still result in an increase of 8\% in the number of casualties and an increase of 6\% in the number of cartel members. Thus, doubling arrests will still translate to a rise in violence compared to the 2022 levels (Figure \ref{PolicyEffects}). Incapacitation is not an optimal strategy for fighting cartels. 

\begin{figure}[h] \centering
\begin{center}
\includegraphics[width = 0.6\linewidth]{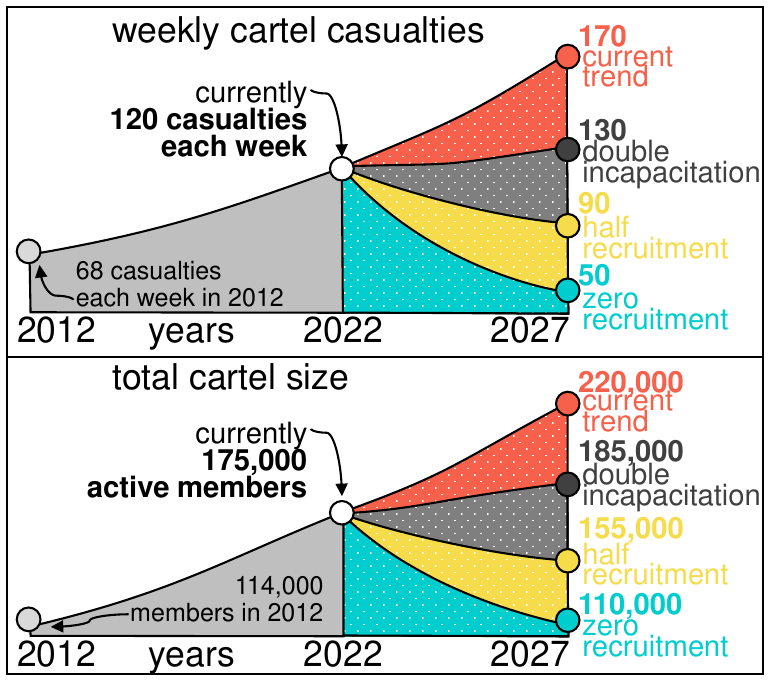}
\end{center}
\caption{Weekly cartel-related deaths (top) and cartel size (bottom) if trends continue, incapacitation doubles, recruitment is reduced by half and recruitment is reduced to zero. Estimates for 2027 are obtained by keeping the 2022 estimates and adjusting the corresponding values of incapacitation or recruitment. } \label{PolicyEffects}
\end{figure}

Conversely, decreasing the cartel's ability to recruit by half will reduce the weekly casualties by 2027 by 25\% and cartel size by 11\%. Mathematically, a preventive strategy is far more successful than a traditional reactive strategy, producing a series of positive spillover effects, including offering alternative pathways to individuals at risk, besides mere violence reduction. However, the cartel population is so large at this point that, even in the hypothetical scenario where recruitment drops to zero, it would take three years to return to the -- already high -- levels of violence observed in 2012. This further calls for rapid and timely large-scale initiatives to reduce recruitment in the country. 

Cartels have a critical equilibrium where their recruitment compensates for their losses, maintaining a stable size. Yet, if the recruitment rate of a cartel is 10\% above its equilibrium, the incapacitation rate has to increase more than 21\% to dismantle it (Supplementary information D). If cartels manage even a slight increase in the recruitment rate, it would need to be compensated by a much more significant increase in the incapacitation rate.

We also assess the effects of two additional ancillary policy scenarios. The first one is designed to alter the type of conflict between cartels by varying the conflict of their actions (for instance, pushing for a \textit{narcopeace}), the second one, targeted at modifying cartels' saturation levels, makes cartels more fragmented (Supplementary information D). Neither of the two strategies outperforms the positive effects that a reduction in recruitment could produce. Decreasing the conflict by 20\% reduces the number of casualties by 8.7\% while increasing saturation by 20\% lowers the number of homicides between cartel members by 5.4\% (Supplementary information D).

}

%%%% PEACE IS NOT FORESEEN
{
In light of the current estimated circumstances, the growth of cartels' size is mostly impeded by the conflict existing among organisations rather than the ability of the state to successfully reduce the levels of violence in Mexico. Our model indicates that, given the current number of cartel members, their recruitment capacity and the limited effect of state forces, the country will evolve into having even more conflicting cartels, with big cartels fragmenting into smaller criminal groups and a large population of individuals at risk of being involved in cartels' violence (Supplementary information E). Cartels will remain salient and impactful in Mexican society for years. 
}

\section{Conclusion}
%%% cartels are relevant
{
For the last 15 years, Mexico has suffered from staggering levels of violence. Most of the violence has been perpetrated by cartels fighting against each other \cite{CongressionalResearchServiceMexicoOrganizedCrime2020}. The impact of cartels on society is pervasive, and notwithstanding the significant human losses caused by such conflicts, cartels have been able to maintain their power without vanishing. Despite the economic, political and societal relevance of cartels, however, and despite a rich literature aiming at studying them, we lack basic information on the size of the population of individuals affiliated with cartels, as well as knowledge on the impact of different policies seeking to curb their power. The lack of reliable data on cartels' size, cartel-related homicides and incarcerations represents the motivating goal of this work and its inherent limitation. We specifically sought to shed light on the black box of cartel dynamics in Mexico, relying on a complex system approach and taking several measures to ensure that the outcomes of our study are robust to alternative modelling choices and assumptions. 
}

%%% reintroduce model
{
Although most cartel-related activities remain undercover, and without rich, fine-grained, reliable data on their characteristics, we have leveraged four publicly available data sources that can be used to infer their size and their behavioural dynamics. In particular, we build on data on homicides, missing persons and arrests over the last ten years in the country and on data on alliances and rivalries among Mexican cartels available for the year 2020. Data on homicides, missing persons and arrests show that human losses have increased over the period under consideration. The network data on alliances and rivalries offer a rich picture capturing the complex nature of conflicts among cartels across different states in Mexico, unfolding hierarchies of power and resources. Exploiting this information, we represent cartels' behaviours through a system of 150 differential equations (as many as the number of cartels included in our data), in which we model the four sources of variation in cartel size, namely recruitment, incapacitation from the state, conflict and saturation. The model is a simplified description of the mechanics of cartel dynamics. We mathematically demonstrate that the ability of cartels to compensate for the losses suffered due to high levels of violence is driven by their ability to recruit a new workforce to remain operative. 
}

%%% re-state quanty results
{
We estimate that by 2022, the number of cartel-affiliated individuals in Mexico was 175,000 units, making cartels the fifth largest employer in the country. Each year approximately 19,000 people are recruited by cartels, many of whom will end up dead or arrested. Furthermore, we highlight how the current size of the cartel population, coupled with the existing levels of violence among cartels, will translate into high levels of violence in the country for years. Even in the hypothetical scenario of a  50\% substantial drop in recruitment, it would take ten years to take back the levels of violence to the ones experienced by the country in 2012.
}

%%% compare four policies
{
In an attempt to offer hints on possible strategies to curb cartels' violence in Mexico, we also assessed the effectiveness of two main scenarios: proactive, intended to prevent recruitment, and reactive, designed to increase incapacitation through arrests. We use our estimate within five years to compare different strategies. Given the current size of cartels, violence will keep increasing at devastating levels. If current levels of incapacitation are doubled, some violence will be contained, but still, we would expect an increase of 8\% in the weekly casualties. Conversely, reducing recruitment by half leads to a decrease in homicides of 25\%. We also test the effect of two ancillary scenarios: reducing the conflict by pushing for cartel agreement and fragmentation, intended to decrease cartels' power through internal fights (Supplementary information D). Results show that the proactive strategy remains substantially more effective in reducing violence in the country. Tackling recruitment has a triple effect: first, it lowers the number of cartel members, reducing the violence it can create by having fewer killers. Second, it lowers the number of targets, so fewer people are vulnerable to suffering more violence. And third, it reduces the cartel's capacity for future recruitment.
}

%%% implications 
{
Our study has several relevant implications for broader research and policy. Research-wise, our modelling approach and the robustness of the results suggest how complex systems can aid the study of criminal phenomena, especially when imperfect proxies of behaviours are the only quantitative source of information available, which can apply to other types of criminal organisations like mafias or terrorist groups. Policy-wise, the outcomes of this work call for profoundly reformative approaches to how cartels are countered. More than 1.7 million people in Latin America are currently incarcerated, and adding more people to saturated jails will not solve the insecurity problem \cite{PrisonBrief}. Many initiatives to counter organised crime (in Mexico and abroad) aim to increase incapacitation through incarceration. Here we indicate how increasing incapacitation substantially may not positively impact the levels of violence, increasing it rather than reducing it, given the current violent inertia in the country. Contrarily, policies oriented toward preventing further recruitment will have longer-lasting beneficial effects. The ability of Mexican authorities to investigate and prosecute crimes, especially severe ones like homicides, is often questioned, with proposals and calls to adopt better police-related measures intended to structurally curb violence \cite{Felbab-BrownCrimeanticrimepolicies2022}. While enforcing justice fairly and effectively is necessary for creating a more peaceful and just society, deploying more effective investigation and prosecution strategies is not the only available measure. Concentrating on incapacitation through traditional forms of reactive policing will not reduce the levels of violence in the country in the expected ways. Conversely, targeting recruitment, preventing people from joining cartels, would lead to extensive reductions in violence. 
}

%%%% reduce recruitment
{
Besides causing sizeable violence reductions, preventing recruitment has additional positive spillover effects. It provides alternative pathways to individuals who would instead become targets of cartel violence or state incapacitation efforts. Preventing recruitment would keep individuals at risk away from the prison and criminal justice systems, thus blocking the array of detrimental effects incarceration has on people \cite{HaneyPrisonEffectsEra2012, MassogliaIncarcerationHealth2015}. Yet, we recognise that reducing recruitment represents a challenge. It requires structural, and yet tailored, exhaustive efforts at the state and local levels to offer male youth and young adults --- who have increasingly become a primary target of cartels' recruitment \cite{BurnettMexicanDrugCartels2009, BreckinHalconesForgottenChildren2019} --- educational and professional opportunities that outweigh the short-term benefits offered by cartels' attractive strength \cite{jones2018strategic, levitt2000economic}. Such a challenge implies cultural, political, economic, and welfare reforms that disrupt the channels through which cartels can exert their persuasive (and often violent) power, especially in areas where cartels benefit from high support from the population \cite{MurphyFollowingpoppytrail2020}.
}

\section{Methods}

\subsection{A dynamic model of cartel size}

%%% introduce diff equation with recruitment
{
Let $C_i(t) \geq 0$ be the size of cartel $i$, with $i = 1, 2, \dots, \kappa$ and let $\mathbf{C}(t) = (C_1(t), C_2(t), \dots, C_\kappa(t))$ be the number of members of each cartel at time $t$ measured in weeks. Let $C(t) = \sum_{i = 1}^\kappa C_i(t)$ be the combined number of cartel members. The size of each group changes due to saturation, recruitment, incapacitation, and conflict with other cartels. We combine its impact on a system of differential equations. Let $\rho \geq 0$ be the recruitment rate, so the cartel increases its size instantly according to $\rho C_i$. Thus, after a (small) time interval $h$, the cartel size is $C_i(t+h) = C_i(t) + h \rho C_i(t)$. Taking the limit as $h$ tends to zero, we obtain that $\dot{C}_i =  \rho C_i$. For some initial size $C_i(0)$ the differential equation gives $C_i(t) = C_i(0) \exp ( \rho t)$. Thus, considering only recruitment, we obtain that all cartels should grow exponentially \cite{caulkins2008optimizing}. But cartels then face saturation, incapacitation and conflict with others, preventing them from growing indefinitely. 
}

%%%% incapacitation and saturation model
{
The incapacitation captures state forces' impact on cartels, with roughly 110 incapacitations due to federal crimes each week in 2020 \cite{inegiPenitenciario}. We assume that state forces have a fixed capacity, so more cartel members reduce the probability that one of them is incapacitated \cite{SacerdoteCrimeInteractions}. Thus, the size of the cartel decreases instantly by $\eta C_i/C$, where $C = \sum_i C_i$ is the combined cartel size and $\eta = 110$ represents the weekly incapacitation rate. Then, we assume that large groups struggle to maintain a stable and organised structure. Thus, due to internal instability, groups decrease their size instantly by $\omega C_i^2$ for some small value of $\omega > 0$, known as the saturation rate. With very small values of $\omega$ compared to recruitment, the impact of saturation is significant only for large groups. This effect mathematically prevents cartels from becoming infinitely large, hence putting a limit on cartel size \cite{caulkins2009optimal}. }

%%%% conflict
{
Finally, the impact of conflict depends on the interactions between cartels $i$ and $j$, so the matrix $S_{ij}$ is defined using the following properties of the network. The \emph{rivalry} $R_{ij} \geq 0$ between $i$ and $j$ is the number of state across which groups $i$ and $j$ fights, with $R_{ij} = 0$ if cartels $i$ and $j$ do not fight. The \emph{strength} of a cartel $A_i$ is the number of alliances between the pairs of groups across Mexican states (Supplementary information B). The size of a cartel is negatively affected by the number of fights, but the impact is smaller with higher strength. The impact of conflict between two cartels, $i$ and $j$, is modelled according to the number of killers of the attacking group. Let $H_j(t)$ be the number of killers of cartel $j$, so cartel $i$ suffers instant casualties according to $\theta_H C_i H_j$, where $\theta_H \geq 0$ is the deadly rate of conflict and the proportion of killers within cartels. We assume that casualties are proportional to the number of killers of each group and that all cartels have the same proportion of killers among their members. Combined, the immediate impact on group $i$ related to its conflict with cartel $j$ is proportional to 
\begin{equation}
 - \theta C_i  C_j S_{ij} \text{, where } S_{ij} = \frac{R_{ij} + \epsilon}{A_i + 1},
\end{equation}
so the impact depends on the rivalry between $i$ and $j$ and the strength of $i$. The small value of $\epsilon$ is added so that all groups have some (minor) friction with others. The impact of conflict by adding all cartels is $\theta C_i \sum_{j\neq i} C_j S_{ij}$. Other models of violence between groups look at the impact of risk aversion of its members, the arrest probability, the internal structure of the group or its mobility patterns \cite{chuang2019mathematical, epstein2002modeling, prieto2020uncovering}, or they look at the frequency or severity of their events \cite{enders2000transnational, clauset2010generalized}. Here, we are interested in the cartels' size and recruitment process.
}

%%%% master eq
{
Combining the instant recruitment, incapacitation, saturation and conflict, we obtain equation \ref{eq:MasterEquation}. The coupled system of differential equations gives the rate of change in the size of each cartel. The instant casualties $d(t)$ follow
\begin{equation}
d(t) =  \theta \sum_{i=1}^{\kappa} \sum_{j \neq i}^{\kappa} C_i C_j S_{ij} = \theta \mathbf{C}^{\top}\mathbf{S}\mathbf{C},
\end{equation}
where $\mathbf{S}$ is the \textit{conflict} matrix with entries $S_{ij}$.
}

%%% IVP
{
It has been observed that the size of organisations follows a heavy-tailed distribution, meaning that many employees work in a small number of firms \cite{mansfield1962entry}. Based on that principle, we model the initial size of cartels as a power-law distribution. Sorting cartels in a decreasing order by their strength $A_i$, we assume that for time $t=0$ the initial cartel size $C_i(0) \sim Po(C_0 i^\beta/\sum_{i = 1}^\kappa i^{\beta})$, where $C_0 = \sum_{i = 1}^\kappa C_i(0)$ is the initial number of cartel members. We assume that $\beta = -1$ and that there are $C_0$ initial members of all cartels combined. The expected number of members of cartel $i$ is $C_0 i^{\beta}/\sum_{i = 1}^\kappa i^{\beta}$. This expression enables us to fix the initial number of all cartel members $C_0$ across the whole country and obtain a reasonable distribution of each cartel's initial number of members. 

 Given some initial size for each cartel $C_i(0)$, its conflict matrix $\mathbf{S}$ and values of the parameters of the recruitment $\rho$, the incapacitation $\eta$, the conflict $\theta$ and the saturation $\omega$ we can model what happens to each cartel at time $\tau >0$. We obtain the solution for the coupled system of differential equations with an initial value using \cite{citeDeSolve, citeR}.
}

\subsection{Model calibration}

%%%% parameter estimation
{
The model depends on a set of parameters, including recruitment, incapacitation, saturation and conflict rates, and also, the initial number of cartel members $C_i(0)$. A set of parameters $\mathcal{P}$ produces a given number of casualties $L_j(\mathcal{P})$ for weeks $j \in J$. We approximate the weekly observed number of incapacitations and casualties reported in Mexico. Thus, we minimise the error between our model's predictions and what was observed, so we can consider the parameter estimation as an optimisation problem. For a set of weeks $J$, the squared difference is defined as $E(\mathcal{P}) = \sum_{j \in J} (L_j(\mathcal{P}) - L)^2$. We keep the values of $\mathcal{P}^\star$ that minimise $E(\mathcal{P}^\star)$. In principle, we ask how cartels have managed to kill each other at such high levels and for such a long period. The parameters that we keep are such that any other combination of cartel members and recruitment results in more (or fewer) casualties and incapacitations than observed in Mexico (Supplementary information C).
}

%%% numbers 2021/01/03
{
Not all incapacitations and not all casualties are related to cartels, so we take an homogeneous fraction of each time series and use them for parameter estimation (Supplementary information C). We then vary this fraction to analyse the sensitivity of keeping only a fraction of the casualties and incapacitations. We obtain that the number of criminals in 2012 was $C_{2012} = 115,000$, but cartels have increased their size, so $C_{2022} = 175,000$. Cartels recruit $\rho^\star = 2.52$ people per thousand cartel members. Thus, if one week a cartel has 1000 members, they will recruit 2.52 members. Cartels recruited 260 people weekly in 2012 and nearly 375 people by 2022. Cartels had 68 casualties each week in 2012 and nearly 120 by 2022. 
}

%%%% results from parameter estimation
{
We obtain that the likely number of criminals in 2022 is $C_{2022}^\star = 175,000$, between 160,000 and 185,000 members. The biggest cartel is CJNG, with an estimated 31,000 people (between 28,600 and 33,100 members). Also, we detected the presence of many small cartels with roughly 200 members in 2022. 
}

To compare different policies, we take our 2022 estimate for the size of each cartel, and the recruitment, incapacitation, saturation and conflict parameters. We then model the next ten years of conflict between cartels (between 2022 and 2031). We consider different scenarios. First, with the same estimate for the set of all parameters. Then, we compare if incapacitation increases, or if recruitment is reduced to half or even zero. Results show that even if recruitment goes to zero, it will take three years to go back to the 2012 levels and an additional five years to reduce that violence by half.

\subsection{Model interpretation}

%%%% model interpretation one cartel
{
The model enables us to understand the qualitative behaviour of the system when the recruitment of cartels or the incapacitation rate is changed. Equation \ref{eq:MasterEquation} depends on some parameters that alter the system's behaviour. In its most simple expression, we can consider a single cartel and the impact of incapacitation. With a low recruitment rate, the cartel will eventually vanish, so $C_1 = 0$ forms the unique stable point. However, with a higher recruitment rate (with $\rho > 2\sqrt{\eta \omega}$), a new stable equilibrium is formed, with an increased number of criminals. In this new equilibrium, the cartel compensates for the impact of incapacitation by recruiting new people (Supplementary information E). Under this scenario, if the cartel recruitment is $1+\epsilon$ times above the critical threshold, the incapacitation rate needs to increase by a factor more significant than $(1+\epsilon)^2$ to compensate. 
}

%%%% model interpretation two cartels with figure. 
{
We also observe that the dynamic changes according to the recruitment rate for two conflicting cartels. For small recruitment, incapacitation is enough to dismantle any cartel. However, for a sufficiently large recruitment rate, large cartels prevail (Figure \ref{TwoDimensionPhasePlane}). Given the initial number of cartel members $C_1(0)$ and $C_2(0)$, will the country evolve into a system with no criminals ($C_1 = C_2 = 0$), or will it converge to any of the cartels overtaking the capacity of the police? The \emph{basin of attraction} of the attractor nodes is the set of the initial number of members that eventually converge to that node \cite{strogatz2018, caulkins2009optimal}. Starting from any point in the $C_1(0), C_2(0)$ distribution, we look for the equilibrium point observed in the dynamics after a long time $t$. We distinguish three significant nodes in the system: one, which we call ``peace'', with no cartel members, so $O = (0,0)$, $W_1$ and $W_2$, representing states where cartel 1 or cartel 2 dominate, respectively (Figure \ref{TwoDimensionPhasePlane}). For a low recruitment rate, $\rho$, the system will eventually go to peace, regardless of the initial number of criminals. The system does not continuously evolve into a peaceful state for a higher recruitment rate. It goes to peace if the number of cartel members is small enough, but also with many cartel members fighting against the others. Here, the country becomes peaceful as a result of conflicting cartels. However, for an even higher recruitment rate, the conflict between cartels is no longer enough to create a peaceful state. Only for a very small number of cartel members the power of incapacitation is sufficient to control them. However, the system becomes dominated by cartels for a sufficiently large number of cartel members. 

\begin{figure}[h] \centering
\begin{center}
\includegraphics[width = 0.7\linewidth]{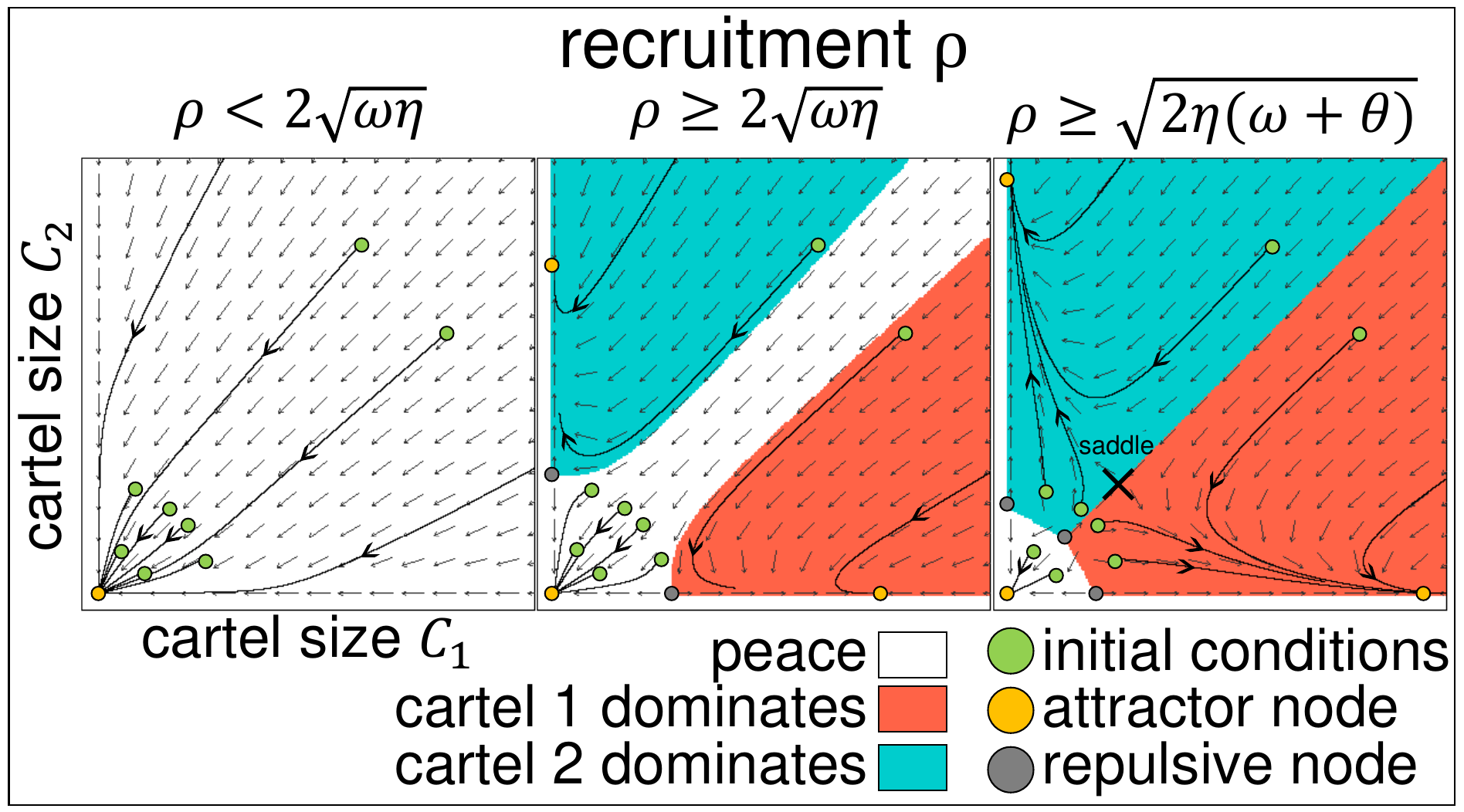}
\end{center}
\caption{Basin of attraction observed for different values of the initial number of cartel members $(C_1, C_2)$ in the horizontal and vertical axis for a varying recruitment rate. The cartels on the left have a low recruitment rate $\rho$, so the system moves away from peace. The cartels in the middle have medium recruitment, and those on the right have high recruitment. The arrows indicate the system's instant movement of the cartel sizes and some trajectories, depending on their initial conditions. } \label{TwoDimensionPhasePlane}
\end{figure}
}

%%%% peace is not always the end state
{
A relevant aspect that occurs with two or more cartels is the impact of conflict. It is desirable, although perhaps unrealistic, for all cartels to vanish. There is a range where many cartel members and a large recruitment rate are driven back to peace, even if cartels exceed what incapacitation can manage. Here, cartels cannot outnumber the state forces, so they cannot avoid vanishing after some period. However, with many cartel members, there is a transition to a perpetual violent state. State forces can no longer contain all cartels and their recruitment. With a sufficiently high cartel recruitment rate, the country does not converge to a peaceful state. Instead, cartels fighting each other becomes pervasive, as observed in most Latin American countries.
}

\appendix

\setcounter{equation}{0}
\setcounter{section}{0}
\setcounter{figure}{0}
\setcounter{table}{0}
\makeatletter
\renewcommand{\theequation}{S\arabic{equation}}
\renewcommand{\thefigure}{S\arabic{figure}}
\renewcommand{\thetable}{S\arabic{table}}

\section{Supplementary information}

\subsection{Data for the fluctuations of violence and incapacitations}  %%%% SUPPLEMENTARY A 

We consider three data sources to model why a cartel loses members: homicides, missing persons and arrests. According to the National Institute of Geography and Statistics in Mexico (INEGI), fewer than 9,000 murders happened in 2007, but nearly 36,000 murders in 2021 \cite{InegiMortalidad}. Thus, the number of homicides in the country increased by more than 300\%, with the population only increasing by 19\%. Roughly 90\% of the victims of homicide victims in Mexico are males, and half are between 20 and 40 years old. The number of people missing and yet to be found increased from less than 11,000 reports in 2012 to more than 22,000 by 2021 \cite{RNPDNO}. Finally, according to the Mexican prison census, roughly 108,000 people were imprisoned each year between 2012 and 2021 \cite{inegiPenitenciario}.

The data correspond to yearly observations between 2012 and 2021. We take the annual reports of homicides, missing persons and arrests and interpolate them into weekly data by applying a cubic spline interpolation \cite{citeR}. The procedure transforms the ten yearly observations into 520 weeks with a smooth transition between one week and the next. The weekly homicides and missing persons yet to be found are combined into the \textit{casualties} time series $T_t$, with $t = 1, 2, \dots, 520$. The weekly number of arrests gives the \textit{incapacitation} time series $I_t$.

Between 660 and 1170 people were killed or reported as missing and yet to be found each week, and between 1700 and 2600 people were imprisoned each week. Thus, we get that $T_t \in [660, 1170]$ and that $I_t \in [1700, 2600]$. However, not all homicide casualties are directly linked to cartel-cartel conflict, many of which are a byproduct of their dispute. Also, most incapacitations are not related to cartels either. Here we assume that a fraction $f \in [0,1]$ of those casualties are directly linked to cartels and that a small fraction $g \in [0,1]$ of the incapacitations are referred to cartel members. We take $f$ and $g$ as model parameters and vary their value to obtain estimates and intervals of the cartel size and their impact (Supplementary information C).

\subsection{Cartel rivalries and alliances} %%%% supplementary B DONE

Data related to cartels in Mexico in 2020 was obtained from open sources, including national and local newspapers and narco blogs and was compiled by CentroGeo, GeoInt and DataLab, part of Consejo Nacional de Ciencia y Tecnología (Spanish for National Council of Science and Technology; abbreviated CONACYT), Mexico's governmental entity in charge of the promotion of scientific and technological activities \cite{BacrimMexico}. Rivalries between cartels are often verified across many sources. The open-access data identified 150 active cartels. Some are national-wide structures, such as Cartel Jalisco Nueva Generación, but others are local gangs. The data has some limitations. Some cartels might not be detected by looking at internet sources, and some alliances and rivalries might also not be seen. Also, data is only available for 2020, so it is impossible to trace cartel activities throughout the years. However, given the sources and the impact of criminal activities in Mexico, we posit that unknown cartels are most likely small local gangs rather than big national organisations. For sensitivity analysis, we evaluate the effect of assuming that the data has 10\% cartels missing and 10\% rivalries missing (Supplementary information F).

The data identifies rivalries between cartels across different states (or 32 provinces) in Mexico. For example, Cártel Jalisco Nueva Generación (CJNG) and La Nueva Familia Michoacana have been detected to fight across 13 states, so we assume that their conflict is more lethal than cartels that only fight in a single state. A weighted network is constructed where a node represents each cartel and edges between $i$ and $j$ represent that cartels have a conflict. The edge weight $W_{ij} > 0$ is the number of states of their dispute. There are 92 conflicting pairs of cartels with a total weight of 179. On average, when two cartels fight, they do it across $179/92 \approx 2$ states. The largest cartel (CJNG) has 77 state-cartel conflicts, reflecting its national presence and high relevance. Also, for 79 cartels (so 53\% of the detected cartels), there are no conflicting organisations, so they are primarily localised groups. We observe that the 28 most prevalent and biggest cartels (19\%) have 80\% of the state conflicts.

Similarly, we construct alliances between cartels across different states in Mexico. We get 91 associations between pairs of cartels with a total weight of 163. Thus, in 2020 there were 163 times when two cartels were allies in different states in Mexico. Although the COVID-19 pandemic could have changed how some cartels fight or cooperate with others, we do not observe any decrease in the number of intentional homicides in Mexico between 2019 (with 34,715 homicides) and 2020 (with 34,562 homicides) according to the Mexican Government \cite{InegiMortalidad}. From all pairwise interactions between cartels ($150 \times 149 / 2 = 11,175$, we get that 24\% of them produce 80\% of the cartel fatalities. Although the data comprises information related to 150 cartels, only the conflict between the most prominent 50 cartels represents more than half of the fatalities related to cartels.

\begin{figure}[h] \centering
\begin{center}
\includegraphics[width =0.8\linewidth]{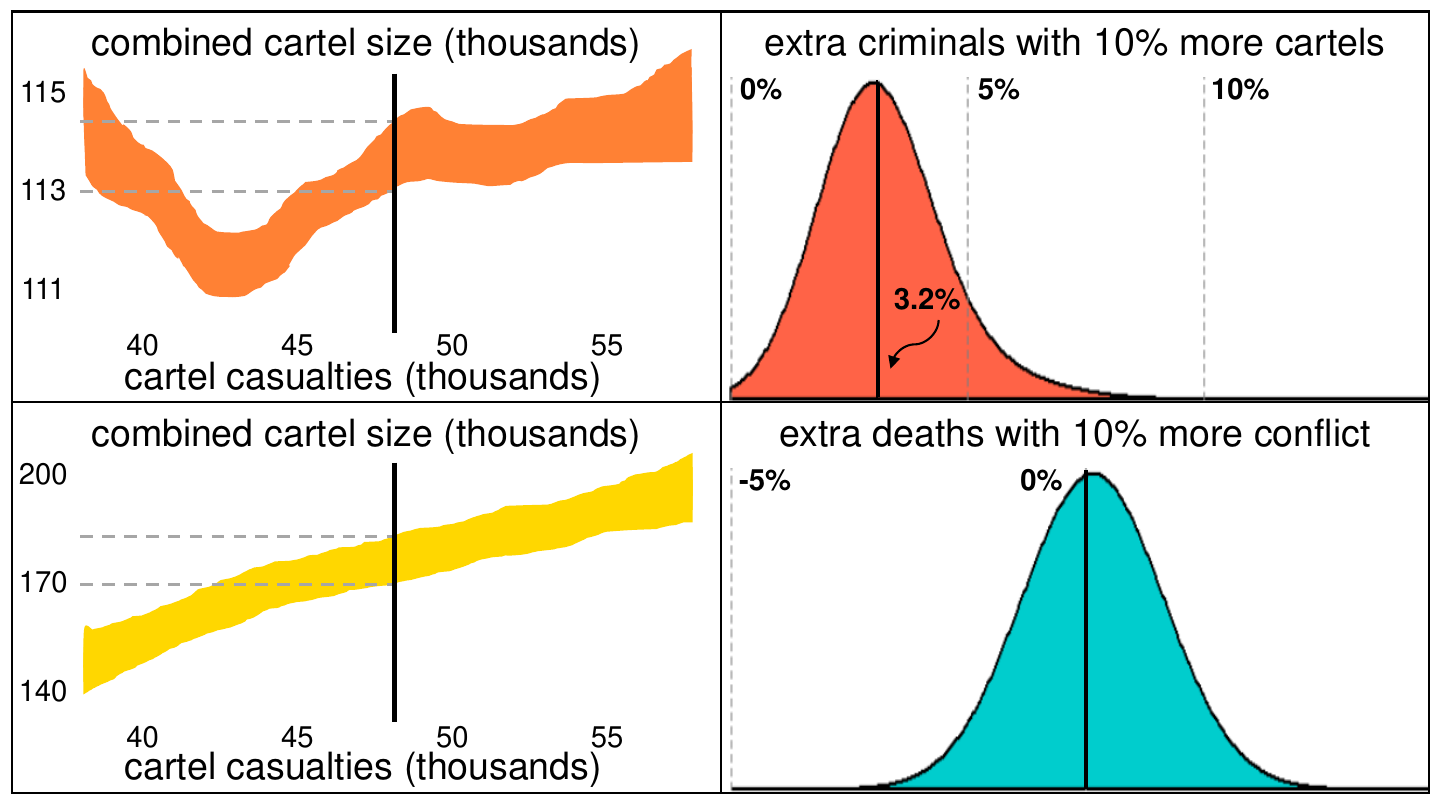}
\end{center}
\caption{Uncertainty intervals obtained for the initial cartel size (2012) and the size ten years later (left). We vary the number of casualties considered (horizontal axis) and obtain different values for the initial and subsequent cartel size. Impact of adding 10\% more cartels in the number of active criminals (top right) and impact of adding 10\% more rivalries in the network (bottom right). } \label{ImpactUncertainty}
\end{figure}

We acknowledge two sources of uncertainty regarding the cartels and the alliances and rivalries networks. The first is related to cartels that are not observed by our data. That means there could be some cartel that does not appear in national and local newspapers and hence cannot be included in the data. The second pertains relationships (either alliances or rivalries) that are not captured. To measure the impact of unknown cartels, we assess a scenario with 10\% more cartels, so instead of 150 active organisations, we consider 165 cartels (Figure \ref{ImpactUncertainty}). We assume that the probability that a cartel is unknown is inversely proportional to its size, so larger cartels are more challenging to maintain undercover. To keep a reasonable cartel size, we randomly take 15 known cartels and assume that the unknown cartels have a size equal to the sampled cartels. We repeat the same procedure 1,000 times and measure the combined number of cartel members. Results show that if there are 10\% more cartels in Mexico than the known ones in 2020, there would be between 1.3 and 4.3\% more cartel members. On average, with 10\% more cartels, there would be 3.2\% more units than the estimated ones.

Concerning possibly undetected relationships, we randomly add 10\% rivalries to the network, so if they were undetected, we consider its impact (Figure \ref{ImpactUncertainty}). We repeat the same procedure 1,000 times and obtain that the total number of casualties in ten years increases, on average, less than 1\% when adding 10\% more state rivalries, demonstrating the robustness of the original data. We follow the same procedure for cartel alliances. There are 163 cartel alliances across different states, so we randomly add 16 partnerships to the network and measure the simulated number of casualties after ten years. Results show that with 10\% more alliances, the fatalities would drop between -1.1 and -0.8\%. Even in this case, the impact of possibly missing relationships is tiny.

In general, although the data has some limitations, the impact of its uncertainty remains within reason. If we are unaware of 10\% of the cartels, their rivalries or their alliances, the impact would be, at most, 5\% of the overall dimension of violence. 

\subsection{Cartel size and parameter estimation} %%%% supplementary C done

We establish the initial size of cartels with a power-law distribution. We assume that for time $t=0$ the initial cartel size $C_i(0) \sim Po(C_0 i^\beta/\sum_{i = 1}^\kappa i^{\beta})$, where $C_0 = \sum_{i = 1}^\kappa C_i(0)$ is the initial number of cartel members, with $\beta = -1$ and $C_0$ the initial number of members of all cartels combined. The expected number of members of cartel $i$ is $C_0 i^{\beta}/\sum_{i = 1}^\kappa i^{\beta}$. This expression enables us to fix the initial number of all cartel members $C_0$ across the whole country and obtain a reasonable distribution of each cartel's initial number of members. We consider recruitment, incapacitation, conflict and saturation between cartels and obtain
\begin{equation} \label{MasterEquationSupplementary}
\dot{C}_i = \underbrace{\rho C_i}_\text{recruitment} - \underbrace{\eta \frac{C_i}{C}}_\text{incapacitation} - \underbrace{\theta \sum_{j \neq i}^{\kappa} C_i C_j S_{ij}}_\text{conflict} - \underbrace{\omega C_i^2}_\text{saturation}. 
\end{equation}

The weekly recruitment in the country $r(t)$ obeys
\begin{equation}
r(t)  = \rho \sum_{i=1}^{\kappa} C_i = \rho C,
\end{equation}
the weekly casualties $d(t)$ follow
\begin{equation}
d(t) =  \theta \mathbf{C}^{\top}\mathbf{S}\mathbf{C},
\end{equation}
the weekly incapacitation $i(t)$ is
\begin{equation}
i(t) = \eta \sum_{i = 1}^\kappa \frac{C_i}{C} = \eta \text{, if $C >0$ and 0 otherwise},
\end{equation}
and the weekly saturation $q(t)$ is 
\begin{equation}
q(t) = \omega \sum_{i = 1}^\kappa C_i^2.
\end{equation}

The instant recruitment and saturation are unknown, but we can infer the incapacitation and casualties, $g I_t$ and $f T_t$, for some $f$ and $g$. For a set of parameters $\mathcal{P} = (\rho, \eta, \theta, \omega)$, for some initial cartel size $C_0$ and for some value of $f$ and $g$, we compute the squared difference, defined as $E(\mathcal{P}) = \sum_{t \in 1, \dots, 520} (f T_t(\mathcal{P}) - d(t))^2 + (g I_t(\mathcal{P}) - i(t))^2$. Departures from the observed values of $g I_t$ and $f T_t$ are less favoured. We keep the values of $\mathcal{P}$ that minimise the function $E(\mathcal{P})$. 

With $f = 0.1$ and $g = 0.05$, we obtain that the number of cartel members in 2012 was $C_{2012} = 115,000$, but cartels have increased their size, so $C_{2022} = 175,000$. Although it is impossible to know with precision the number of cartel members and their recruitment capacity, the number of kills can only be observed under specific circumstances, including a high volume of cartel members and recruitment rate. 

\subsection{Varying saturation and the type of conflict} %%%% supplementary D

\begin{figure}[h] \centering
\begin{center}
\includegraphics[width = 0.8\linewidth]{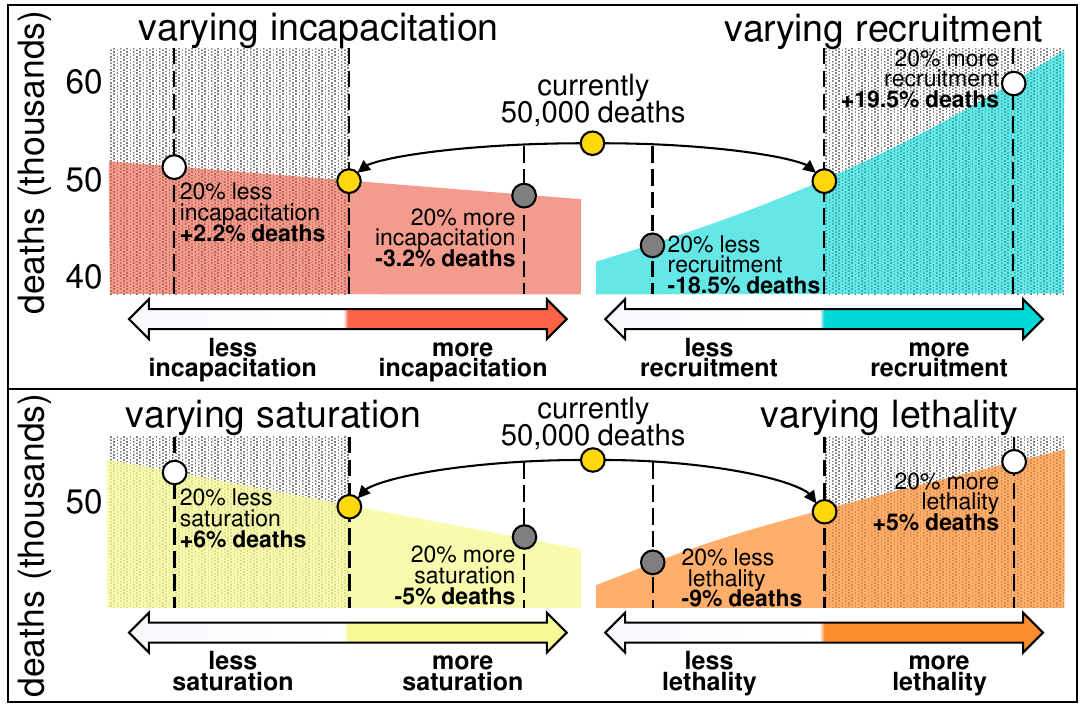}
\end{center}
\caption{Impact of varying the incapacitation, the recruitment, the saturation and the conflict of cartels within a reasonable range and measuring the number of casualties within ten years. } \label{SIFigureFragmentationLethality}
\end{figure}

Varying the type of conflict with cartels having more or less saturation and having more or less conflict alters the casualties within ten years (Figure \ref{SIFigureFragmentationLethality}). If the conflict is more lethal, it increases the number of deaths suffered since cartels compensate for their losses by recruitment. If cartels have more fragmented structures, the number of casualties decreases. Increasing the saturation by 20\%, thus, making it more difficult for cartels to form large organisations, is to reduce the number of deaths by 5.4\%. Reducing the conflict by 20\% decreases the number of casualties by 8.7\%.

\subsection{A dynamical system of cartel size} %%%% supplementary E

The cartel size model in equation \ref{MasterEquationSupplementary} is a dynamical system with $\kappa = 150$ coupled differential equations. We analyse its fixed points (those points in which $\dot{C} = 0$), and the corresponding basin of attraction \cite{strogatz2018}. First, we analyse the scenario with $\kappa = 1$, so only one cartel against the state forces, and then we analyse competition between two cartels, so $\kappa = 2$. For more cartels, the qualitative description of the results is similar.

\subsubsection{One cartel against the police}

\begin{figure}[h] \centering
\begin{center}
\includegraphics[width = 0.8\linewidth]{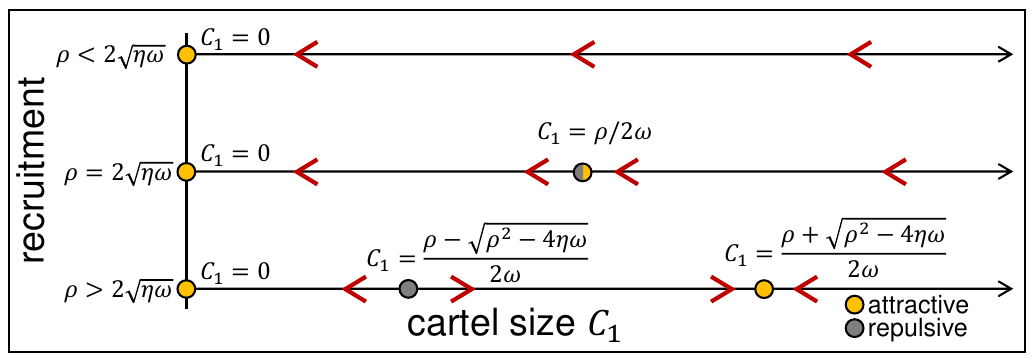}
\end{center}
\caption{Diagram with only one cartel. Its size is the horizontal axis. We vary the recruitment rate from having a small value (top) to a large value (bottom). Fixed points are marked as circles in the diagram, and the dynamic is represented by arrows.} \label{1CartelDiagram}
\end{figure}

Although simple, the system with only one cartel helps us show the impact of the different parts of the model (Figure \ref{1CartelDiagram}). With only one cartel, their size varies with
\begin{equation}
\dot{C}_1 = \rho C_1 - \eta - \omega C_1^2, \text{for $C_1 >0$ and $\dot{C}_1 = 0$ at $C_1 = 0$.}
\end{equation}
The system always has one fixed point of ``peace'', at $O = \{ C_1=0\}$ and, depending on the parameters' values, it could have one or two extra fixed points. If $\rho = 2\sqrt{\eta \omega}$ there is only one extra fixed point at $H = \{ C_1 = \rho/2 \omega \}$. For a higher recruitment rate, we obtain two extra fixed points, one at $A = \{ C_1 = (\rho - \sqrt{\rho^2 -4 \eta \omega})/2\omega \}$ and at $B = \{ C_1 = (\rho + \sqrt{\rho^2 -4 \eta \omega})/2\omega \}$. The fixed point $O = \{ C_1=0\}$ is always an attractive node. For higher recruitment, there is an attractive node on one side and a repulsive one on the other side, where the impact of incapacitation is in equilibrium with recruitment. With an even larger recruitment rate, there are two non-trivial fixed nodes, one repulsive ($A$) and one attractive ($B$).

Suppose that $\rho = 2\sqrt{\eta_0 \omega} \times (1+\epsilon)$, for some incapacitation rate $\eta_0$ and for some value of $\epsilon > 0$. Thus, there are two attractor nodes at $O$ and $B$. If the number of cartel members is at the stable node $B$, then the incapacitation rate has to increase by a factor $(1+\alpha)$ to guarantee that $2 \sqrt{\eta_0 (1+\alpha) \omega} > 2\sqrt{\eta_0 \omega} \times (1+\epsilon)$. We obtain that $1+\alpha > (1+\epsilon)^2$, meaning that if the recruitment rate of a cartel is 10\% above its critical threshold, then the incapacitation has to increase more than 20\% to dismantle the cartel. As has been observed before, we observe two stable nodes: one with low crime and high probability of arrest and the second with high crime with a low probability of arrest \cite{SacerdoteCrimeInteractions, SacerdoteCrimeCities}.

\subsubsection{Two conflicting cartels}

The dynamics between two conflicting cartels enable us to understand the behaviour of the simplified version of the system. With $\kappa = 2$, and $S_{ij} = 1$, we obtain that
\begin{eqnarray*}
\dot{C}_1 &=& C_1 \left(\rho -\frac{\eta}{C_1+C_2} - \theta C_2- \omega C_1 \right), \text{ and}\\
\dot{C}_2 &=& C_2 \left(\rho -\frac{\eta}{C_1+C_2} - \theta C_1- \omega C_2 \right),
\end{eqnarray*}
a system of two coupled differential equations. Equilibrium points are observed when $\dot{C}_1 = \dot{C}_2=0$. We obtain that $\dot{C}_1 = 0 \Longleftrightarrow C_1 = 0$, or if $\omega C_1^2 + (\theta C_2 +  \omega C_2 - \rho) C_1  + \theta C_2^2 -\rho C_2 +\eta = 0$, and similarly for $\dot{C}_2 = 0$.

Without the intervention of the police (with $\eta = 0$ so the incapacitation rate is zero), there are four equilibrium points, at $O = (0,0)$, $S = (\frac{\rho}{\omega + \theta}, \frac{\rho}{\omega + \theta})$, $W_1 = (\rho/\omega, 0)$, and $W_2 = (0, \rho/\omega)$. The point $O$ is an unstable node, meaning that if the initial number of cartel members is slightly greater than zero (any of the cartels or both), then the dynamics will evolve away from $O$. The point $S$ is a saddle, meaning it can be defined in terms of two asymptotic curves, one with $C_1 = C_2$. The other two points, $W_1$ and $W_2$ are stable nodes. Without police, the dynamics can be described as follows: for any non-trivial initial number of cartel members, the point with no criminals is repulsive, so eventually, there will be more criminals in the city. Further, even if the two cartels have any initial growth in size, any initial advantage will become critical. If $C_1(0) = C_2(0)+\epsilon$, then the impact of that initial advantage $\epsilon > 0$ will eventually allow the group $C_1$ to become the dominant group and the dynamics will converge to $W_1$. The cartels do not become even bigger due to saturation. The cartels' terminal size is $\rho / \omega$. The weekly number of kills is $2 \theta C_1 C_2$. Thus, the conflict between two cartels is more lethal when there are more criminals but, more importantly, when they have a relatively similar size.

With a small intervention from the police, there are some changes to the system. For small values of $\eta$, the peace point $O = (0,0)$ becomes an attractor node. Thus, even with a small number of criminals, the system returns to $O$ after some steps. Here, the police can deal with some criminals and reinstate peace. However, for a larger number of criminals, the police are exceeded. Eventually, if $C_1$ or $C_2$ is greater than some threshold and they have the recruitment capacity, they will exceed the capacity of the police. Formally, if $\rho \geq 2\sqrt{\omega \eta}$ and $C_1(0) = \sqrt{\eta/\omega}$ then the cartel has exceeded the capacity of the police and will grow until it reaches a new equilibrium. With police intervention, we can obtain two, four, five or up to six equilibrium points. If $\rho = 2\sqrt{\omega \eta}$ then there are two additional equilibrium points at $P_1 = (\sqrt{\eta/\omega}, 0)$ and at $P_2 = (0, \sqrt{\eta/\omega})$. For a larger recruitment rate, if $\rho > 2\sqrt{\omega \eta}$ then we obtain four equilibrium points:
$Q_1 = ((\rho + \sqrt{\rho^2-4\omega \eta})/2\omega, 0)$ and $R_1 = ((\rho - \sqrt{\rho^2-4\omega \eta})/2\omega, 0)$ and due to symmetry, the same points $Q_2$ and $R_2$ within the other axis. 

%observed when the two parabolas touch at a single point.
If $\rho = \sqrt{\eta(2\omega + 2\theta)}$ there is a fifth equilibrium point at $S = (\rho / (2\omega + 2\theta), \rho / (2\omega + 2\theta))$. Finally, if $\rho > \sqrt{\eta(2\omega + 2\theta)}$ then we obtain two symmetric equilibrium points, $T_1 = (\rho +\sqrt{\rho^2-\eta(2\omega + 2\theta)})/(2\omega + 2\theta), (\rho +\sqrt{\rho^2-\eta(2\omega + 2\theta)})/(2\omega + 2\theta)$ and a second point at $T_2 = (\rho -\sqrt{\rho^2-\eta(2\omega + 2\theta)})/(2\omega + 2\theta), (\rho -\sqrt{\rho^2-\eta(2\omega + 2\theta)})/(2\omega + 2\theta)$.

Having only two conflicting cartels enable us to get two results. First, if at any point in time $\tau$, we observe that $C_1(\tau) > C_2(\tau)$ then $C_1(\tau+t) \geq C_2(\tau+t)$ for all $t>0$, that is, any initial advantage that the cartel has, remains. Also, if at some point $\tau$ we observe that the number of weekly casualties is increasing, compared to the previous week, then the total cartel size $C_1(\tau) + C_2(\tau)$ is also increasing. To see this, we express the number of casualties as $d(\tau) = 2 \theta C_1(\tau) C_2(\tau)$. If both cartels reduce their size at time $\tau + \delta$, then the number of casualties also has to drop. Therefore, we analyse what happens when one cartel grows (with $C_1(\tau + \delta) = C_1(\tau) + \epsilon_1$) and the other one decreases (with $C_2(\tau + \delta) = C_2(\tau) - \epsilon_2$), for some $\epsilon_1>0$ and $\epsilon_2>0$. If the number of casualties increases, then it means that $(C_1(\tau) + \epsilon_1)(C_2(\tau) - \epsilon_2) > C_1(\tau) C_2(\tau)$ from which we get that $C_2(\tau) \epsilon_1 - C_1(\tau)\epsilon_2 - \epsilon_1 \epsilon_2 > 0$, meaning that cartel 1 had a growth of at least $\epsilon_1 > C_1(\tau) \epsilon_2 /(C_2(\tau)- \epsilon_2)$. Thus, comparing the total cartel size, we get that $C_1(\tau+t) + C_2(\tau+t) = C_1(\tau) + \epsilon_1 + C_2(\tau) - \epsilon_2 > C_1(\tau) + C_2(\tau) +  C_1(\tau) \epsilon_2 /(C_2(\tau)- \epsilon_2)  - \epsilon_2$. Since $C_1(\tau) > C_2(\tau)-\epsilon_2$ as any initial advantage remains, then $C_1(\tau) \epsilon_2 /(C_2(\tau)- \epsilon_2)  > \epsilon_2$, meaning that $C_1(\tau) \epsilon_2 /(C_2(\tau)- \epsilon_2) - \epsilon_2 > 0$ and therefore, $C_1(\tau+t) + C_2(\tau+t) > C_1(\tau) + C_2(\tau)$. Therefore, if the number of casualties has a positive trend, the total cartel size also increases, despite suffering a greater loss.

\begin{figure}[h] \centering
\begin{center}
\includegraphics[width = 0.6\linewidth]{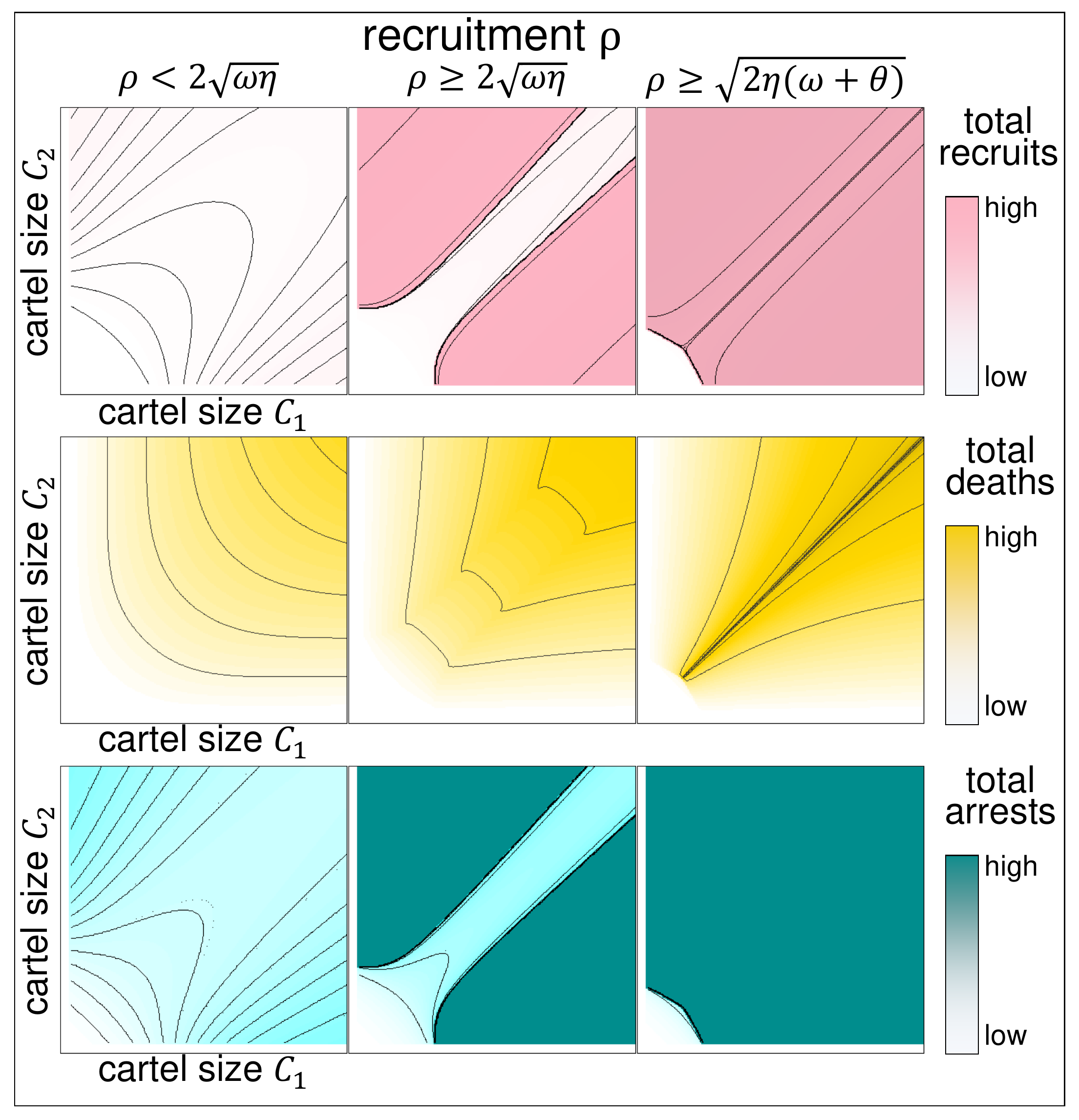}
\end{center}
\caption{Impact of varying the recruitment rate on the total number of recruits (top), the number of deaths (middle) and the total incapacitations (bottom). The analysis represents some initial size of Cartel 1 (horizontal) and Cartel 2 (vertical).  } \label{OtherImpacts}
\end{figure}

\subsection{Cartel size changes (2012-2021)}

\begin{table}[h!]
\caption{Estimated cartel size, recruitment, casualties and incapacitations between 2012 and 2021. Numbers reported in thousands.}
\centering
 \begin{tabular}{c |c c c c | c} 
 \hline
     & Initial& Recruited & Casualties & Incapacitations & Final \\
Year & size & $+$         & $-$        & $-$             & size \\
 \hline
2012 & 115.4 & 13.9 & 3.6 & 5.7 & 120.0 \\
2013 & 120.0 & 14.4 & 3.8 & 5.7 & 124.9 \\
2014 & 124.9 & 15.0 & 4.1 & 5.7 & 130.0 \\
2015 & 130.0 & 15.5 & 4.3 & 5.7 & 135.5 \\
2016 & 135.5 & 16.1 & 4.6 & 5.7 & 141.2 \\
2017 & 141.2 & 16.6 & 4.9 & 5.7 & 147.3 \\
2018 & 147.3 & 17.3 & 5.2 & 5.7 & 153.6 \\
2019 & 153.6 & 17.9 & 5.5 & 5.7 & 160.2 \\
2020 & 160.2 & 18.5 & 5.9 & 5.7 & 167.1 \\
2021 & 167.1 & 19.2 & 6.3 & 5.7 & 175.0 \\
\hline
\end{tabular}
\end{table}

 \subsection{Cartel size in 2022}

\begin{table}[h!]
\caption{Top 20 cartels and their estimated size by 2022}
\centering
 \begin{tabular}{r |c |c| c} 
 \hline
 Group & State rivals & State allies & Size \\ 
 \hline
Cártel Jalisco Nueva Generación (CJNG) & 77 & 55 & 28,764 \\ 
Cártel de Sinaloa & 19 & 34 & 17,825 \\ 
La Nueva Familia Michoacana & 21 & 13 & 10,736 \\ 
Cártel del Noreste & 16 & 10 & 8,992 \\ 
La Unión Tepito & 13 & 9 & 7,561 \\ 
Los Chapitos & 9 & 10 & 6,823 \\ 
Cártel del Golfo & 10 & 7 & 5,556 \\ 
Los Zetas & 10 & 6 & 4,697 \\ 
Guerreros Unidos & 13 & 2 & 3,096 \\ 
Gente Nueva & 4 & 9 & 4,325 \\ 
Zetas Vieja Escuela & 9 & 4 & 3,084 \\ 
Caballeros Templarios & 8 & 4 & 2,686 \\ 
Fuerza Anti-Unión Tepito & 7 & 5 & 2,903 \\ 
Los Rojos & 12 & 0 & 813 \\ 
Cárteles Unidos & 6 & 4 & 2,051 \\ 
Los Mayas & 7 & 3 & 1,824 \\ 
Cártel de Tláhuac & 5 & 4 & 2,050 \\ 
Cártel de Caborca & 4 & 4 & 2,114 \\ 
Los Cabrera & 3 & 4 & 2,016 \\ 
Los Cuinis & 0 & 6 & 2,061 \\ 
Other cartels (130) & 105 & 133 & 55,023 \\ 
\hline
Total & 358 & 326 & 175,000 \\ 
\hline
\end{tabular}
\end{table}

\newpage

\section*{Acknowledgements}

We are grateful for the insightful comments from Lisa Sánchez and Carlos A. Pérez Ricart.

\section*{Competing interests}

The authors declare that they have no competing interests.

\section*{Author's contributions}

\section*{Funding}

The research was funded by the Austrian Federal Ministry for Climate Action, Environment, Energy, Mobility, Innovation and Technology (2021-0.664.668) and the Austrian Federal Ministry of the Interior (2022-0.392.231)

 \newpage

\bibliographystyle{unsrt}

\end{document}